\documentclass[11pt,a4paper,twoside]{article}

\usepackage{graphicx}
\usepackage{fancyhdr,calc,cite,float}
\usepackage{epsfig}
\usepackage[centertags]{amsmath}
\usepackage[mathscr]{eucal}
\usepackage[T1]{fontenc}
\usepackage{amssymb}

\title{\bf The interaction between the projectile and 
the target core in single ionization of helium  
by $100$ MeV/u C$^{6+}$ and $1$ GeV/u U$^{92+}$ ions } 

\author{ A.B. Voitkiv and B. Najjari \\  
Max-Planck-Institut f\"ur Kernphysik \\ 
Saupfercheckweg 1, 69117 Heidelberg, Germany }

\date{\today}  

\begin{document}

\maketitle
 
\begin{abstract} 

We consider single ionization of helium by 
fast highly charged ions. The ionization is treated 
as a three-body problem which involves the 'active' target  
electron, the target core (consisting of the target nucleus 
and the 'passive' target electron) and the projectile-ion. 
We show that, unless the transverse momentum transfer 
in the collision is large, the deviation of the interaction between 
the projectile and the target core from the purely coulomb 
law has a weak effect on the ionization.  
In particular, the account of this deviation does not 
enable one to get an agreement with experimental data  
on the fully differential cross section for 
the electron emission into 
the so called perpendicular plane.  

\end{abstract} 



\section{Introduction} 

Collisions of fast highly charged ions with 
helium represent an interesting example of quantum dynamical 
few-body systems \cite{1}. During the last several years 
single ionization of helium has attracted a lot of attention, 
both experimental and theoretical \cite{he-s-1}-\cite{he-s-9}. 

The most detailed information on the process of ionization is 
obtained by exploring the collision dynamics on the level of  
the fully differential cross section (FDCS). 
The parameter $\eta_p = Z_p/v_p$,
where $Z_p$ is the projectile charge and 
$v_p$ the collision velocity, is commonly used 
to characterize the effective strength of 
the projectile-target interaction. 
At small values of $ \eta_p $ ($ \eta_p \ll 1$), 
where the projectile-target interaction is weak, 
one normally expects that already the first-order 
Born approximation should yield 
a reasonable description of the collisions. 

However, such expectations have been put into question 
by experimental results on single ionization of helium by 
$100$ MeV/u C$^{6+}$ projectiles. At this impact energy 
the collision velocity is quite high ($v_p =58$ a.u.) 
leading to the collision strength parameter $ \eta_p $ 
of just about $0.1$. Nevertheless, large discrepancies  
were found between the experimental data and the first 
Born results for the electron emission outside the scattering plane 
\footnote{The scattering plane is a plane 
spanned by the vectors of the initial and final 
projectile momenta.}. At the same time, for the electrons 
emitted into the scattering plane  
the experimental FDCS was reasonably well reproduced by 
first Born theory \cite{he-s-1}.  

It was suggested in \cite{he-s-2} that 
the discrepancies might have their origin 
in the interaction between 
the projectile and the target core (the target nucleus 
and the passive atomic electron). According 
to the first Born approximation, this interaction does 
not influence the ionization transition amplitude. 
Therefore, a number of theoretical models including, 
in particular, the second Born, the Glauber, 
the continuum-distorted-wave-eikonal-initial-state 
(CDW-EIS) and the symmetric-eikonal approximations  
were applied to study the electron emission. 
All these models go beyond the first Born approximation 
and, in particular, include the interaction 
between the projectile and the target core.       
 
These models predicted noticeable deviations  
in the electron emission pattern from 
the first Born results, especially 
outside the scattering plane \cite{he-s-10}. 
It also should be noted that in 
the case of the ionization by $100$ MeV/u C$^{6+}$ ions 
the predictions of these models concerning the character 
of the deviations and their magnitude 
were quite similar \cite{he-s-10}. 
In particular, according 
to all of these models the main reason  
for the deviations in the FDCS 
from the first Born results in this case is  
the interaction between the projectile 
and the target core.     

These models, however, did not enable one to get 
any better overall agreement with the experiment 
\cite{he-s-2}. Moreover, for the FDCS for the electrons emitted 
into the plane, perpendicular 
to the transverse momentum transfer (the perpendicular plane),   
they all predicted a minimum exactly there where, 
according to the experiment, a maximum should be.  
Since for this plane the first Born approximation 
predicts almost a constant value for the emission 
pattern (see also figure \ref{carbon-3}) 
\footnote{ According to the first Born approximation 
the electron emission pattern should be a constant 
in the plane, which is perpendicular 
to the vector of the total momentum transfer. 
However, under the conditions of the experiment \cite{he-s-2}, 
because of high collision velocity,  
the minimum momentum transfer was very small 
and the transverse part of the momentum transfer 
practically coincided with 
the total momentum transfer.}, 
the disagreement between the experiment and 
theory actually had became even worse.    

It was then suggested \cite{he-s-8} that 
the difference between the experiment and theory 
is not a signature of any principal shortcoming 
of the atomic collision models but is simply 
caused by the thermal motion of the target atoms. 
Because of this motion a detection of the final 
momentum of the recoil ion does not allow one 
to get precise information about the value and 
direction of the momentum transfer in the collision. 
As a result, the orientation of the perpendicular plane 
becomes experimentally not very well defined and the 
measured emission into this plane can contain noticeable 
admixtures of the emission into other planes, 
including 'traces' of the (relatively very large) 
binary peak from the scattering plane.  

The role of possible experimental uncertainties 
was later studied in \cite{he-s-9}.  
It was confirmed that a non-zero 
temperature of the target gas indeed results 
in the appearance of a maximum in the emission spectra 
exactly there where it was observed in the experiment. 
Besides, the authors of \cite{he-s-9} found out that 
the size of the projectile beam also contributes to 
the appearance of the maximum. Nevertheless, 
the above two effects seem to be insufficient 
in order to fully explain the difference between 
the experiment data and theoretical results. 

In another recent paper \cite{he-s-11} 
it was suggested that the reason for the discrepancies  
could lie in the fact that in the theoretical models, 
used so far to describe this process, 
the interaction between the projectile 
and the target core was taken as a pure coulomb 
interaction between two point-like charges:  
$Z_p$ and $1$ for the projectile and target core, 
respectively. Strictly speaking such an approximation 
is only valid for collisions with impact parameters 
larger than the size of a helium atom. 
If a substantial part of the electron emission 
arises in collisions, in which the projectile penetrates  
the target core, the above theoretical models 
might be not able to yield a satisfactory 
description of the ionization process. 
A similar suggestion has been also made 
in \cite{he-s-11a}. 

It is, therefore, one of the main intentions of the present 
paper to attempt to find out whether the account 
of a non-coulomb part of the interaction between 
the projectile and target core 
could bring substantial changes into the form 
of calculated cross sections and, in particular, whether 
this account can improve the description of the experimental data 
on single ionization of helium by $100$ MeV/u C$^{6+}$ 
projectiles. Besides, for a comparison we shall also 
consider single ionization of helium by $1$ GeV/u U$^{92+}$ ions. 
In the latter case the collision velocity is 
even by a factor of $2$ higher ($v_p =120$ a.u.) 
but the parameter $\eta_p =Z_p/v_p$ 
is rather close to $1$ ($\eta_p = 0.77$). 
 
Atomic units are used throughout 
except where otherwise stated. 

\section{Theoretical model} 

We shall treat single ionization of helium  
by considering a three-body model of this process. 
This model involves the active target electron (1), 
the target core (2) and  the projectile (3). 
The projectile and the active electron interact 
via the (relativistic) Coulomb force. The target core consisting of 
the target nucleus and the passive electron is supposed to behave  
in the collision as a rigid body whose internal structure 
is not changed by the collision. 

It is well known that even in collisions 
at relativistic impact velocities 
the majority of electrons emitted from such a light target
like helium has velocities with respect to
the target recoil ion which do not exceed  
a few atomic units.  Therefore, we shall consider the ionization 
using the rest frame of the target nucleus and describe 
the active electron nonrelativistically. 

In this frame the field of the target core 
acting on the active electron and 
the projectile is described by a scalar potential. 
This potential is taken to be the same for 
the active electron and the projectile 
(as well as for the initial and final collision channels) 
and is approximated by 
\begin{eqnarray} 
\Phi(\xi) = \frac{1}{\xi} + \left(1 + \beta \xi \right) %
\frac{ \exp(- \alpha \xi) }{\xi}.    
\label{potential} 
\end{eqnarray}   
Here $\xi$ refers to the distance measured 
from the center of the target core 
(i.e. from the target nucleus) and is given 
in the target frame. In accordance with results 
of \cite{hartree-fock} the values of the parameters 
$\alpha$ and $\beta$ were chosen to be 
$3.36$ a.u. and $1.665$ a.u., respectively. 

The initial state of the colliding system is taken as 
a product of unperturbed states of the projectile and 
target electron multiplied by the eikonal factors. 
These factors reflect the distortion of the initial state 
caused by the coulomb interaction between 
the projectile and the active electron and 
by the interaction between the projectile 
and the target core. The final state is represented by 
a similar product where, however, the eikonal distortion factor 
for the projectile-active electron subsystem is replaced by 
a (relativistic) continuum-distorted-wave factor.   
The initial and final unperturbed states of 
the electron moving in the field of the target core 
are obtained by a numerical solution of 
the corresponding Schr\"odinger equation.  

The main difference between the approach of the present paper 
and our previous attempt \cite{he-s-6}-\cite{he-s-7} 
to address the problem of helium single ionization by very fast 
ions is that the interaction between the projectile and 
the target core, in addition to the coulomb part, 
involves also the short-range interaction   
(see Eq.(\ref{potential})). Besides, an eikonal-like description 
of the projectile-active electron coupling in the final channel 
has now been replaced by the continuum-wave coulomb distortion factor.   
The latter is taken in such a way as to account for  
the necessary relativistic changes appearing 
in the form of the coulomb field generated 
by a very fast moving ion. 
   
The first-order counterpart of the present 
three-body model is obtained if all the distortion factors 
are set to unity. In particular, within the first order approach 
the interaction between the projectile and the target core 
completely drops out from the transition amplitude. 
Besides, the interaction between the projectile 
and the active target electron is reduced just to 
a single-photon exchange. 

\section{Results and discussion}  

\subsection{The fully differential cross section} 
 
Let us now turn to considering the FDCS 
for single ionization of helium, 
$\frac{ d \sigma^+ }{d^2 {\bf Q} d^3 {\bf k}} $.
Here, $ {\bf Q} $ is the transverse part of the
total momentum transfer to the target which 
given by $ {\bf q} = ({\bf Q}, q_{min}) $, where 
$q_{min} = (\varepsilon_k - \varepsilon_0)/v_p$
with $\varepsilon_0$ and $\varepsilon_k$ being the 
initial and final energies of the active electron.   
${\bf k}$ is the momentum of the emitted electron
with respect to the target nucleus.
We shall first consider the ionization caused by 
$100$ MeV/u C$^{6+}$ projectiles. 

\begin{figure}[t]
\vspace{0.25cm}
\begin{center}
\includegraphics[width=0.55\textwidth]{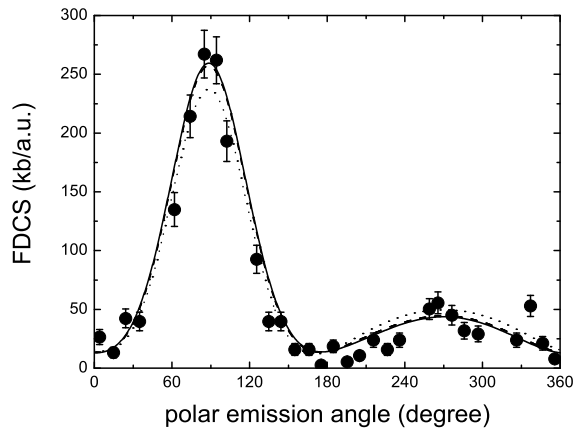}
\end{center}
\caption{ \footnotesize{ The FDCS for single
ionization of helium by $ 100 $ MeV/u C$^{6+}$ 
projectiles plotted as a function of the polar emission angle.
The cross section is given in the scattering plane. 
The electron emission energy $\varepsilon_k = 6.5 $ eV 
and the total momentum transfer $ q = 0.65 $ a.u.. 
Symbols: experimental data from \cite{he-s-1}. 
Solid curve: results obtained by including into the interaction 
between the projectile and the target core 
the full expression (\ref{potential}). 
Dash curve: the projectile-target core interaction 
is modeled by the coulomb part of (\ref{potential}) only. 
Dot curve: the first Born results. }  }
\label{carbon-1} 
\end{figure}

\begin{figure}[t]
\vspace{0.25cm}
\begin{center}
\includegraphics[width=0.55\textwidth]{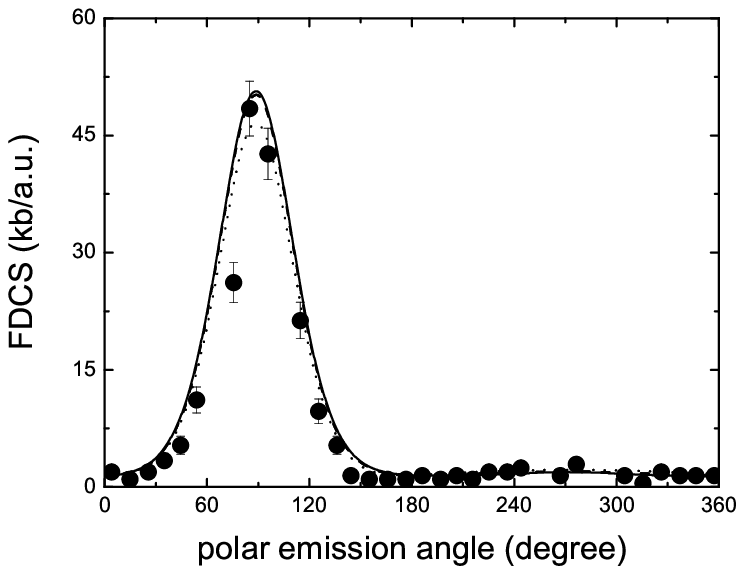}
\end{center}
\caption{ \footnotesize{ Same as in figure \ref{carbon-1} 
but for the electron emission energy of $17.5$ eV and 
the total momentum transfer of $1.43$ a.u.. }}
\label{carbon-2} 
\end{figure}

\begin{figure}[t]
\vspace{0.25cm}
\begin{center}
\includegraphics[width=0.55\textwidth]{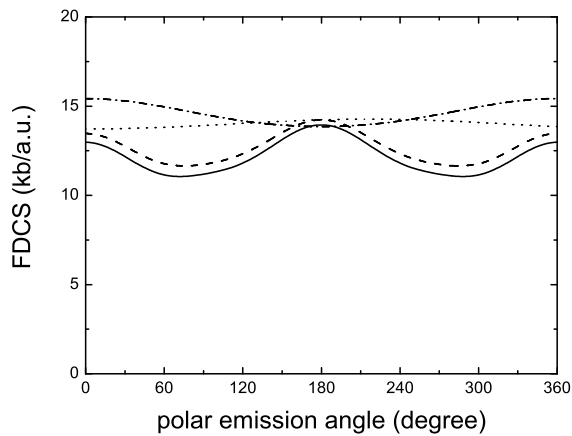}
\end{center}
\caption{ \footnotesize{ Same as in figure \ref{carbon-1} 
but the cross section is given in the perpendicular plane. 
Besides, the dash-dot curve displays results 
of the calculation in which the interaction between the projectile and 
the active electron is described within the rCDW-EIS but  
the projectile-target core interaction is ignored. }  }
\label{carbon-3} 
\end{figure}

In figures \ref{carbon-1}-\ref{carbon-2} we 
show the FDCS for the scattering plane. This plane 
is defined by the condition $\varphi_k=0^0$, 
where the azimuthal angle $\varphi_k$ of the momentum 
${\bf k}$ of the emitted electron is counted from 
the direction of the transverse momentum transfer ${\bf Q}$. 
This cross section is given as a function of the polar angle,
$ \vartheta_{\bf k} = \arccos({\bf k} \cdot {\bf v}_p /kv_p )$,
of the momentum of the emitted electron for fixed emission energy 
and momentum transfer. The results are on an absolute scale. 

We observe in these figures familiar structures:  
at the lower momentum transfer (see figure \ref{carbon-1}) 
the emission pattern clearly exhibits 
the so called binary and recoil peaks; 
at the higher momentum 
(see figure \ref{carbon-2}) the recoil peak practically disappears.  
It is also seen in these figures that, while the inclusion of 
the interaction between the projectile and the target core 
has a visible (although rather weak) 
impact on the calculated cross section, the latter 
is practically insensitive to whether this interaction is 
approximated by merely the coulomb part 
of the scalar potential (\ref{potential}) or by 
the full expression (\ref{potential}).  

The situation becomes somewhat different if we consider 
the emission into the perpendicular plane (see figure \ref{carbon-3}). 
The latter is defined by the condition $\varphi_k=90^0$. 
The emission into this plane is very weak and much more sensitive 
to the details of a theoretical description.   
For this plane the deviations between the results calculated 
in the different approximations become more pronounced. 
In this plane the higher order contributions in the interaction between 
the projectile and the active electron change the shape 
of the cross section. The inclusion of the interaction between the 
projectile and the target core brings further quite noticeable   
changes into the calculated emission pattern. Moreover, in 
the perpendicular plane we also observe that there is a visible 
effect in the emission spectrum caused by 
the short-range part of this interaction.  

However, in contrast to the experimental data showing a pronounced maximum 
at $\vartheta_k \sim 90^0$ and $270^0$ (see e.g. \cite{he-s-9}), 
the inclusion of the interaction between  
the projectile and the target core results in a minimum 
in the calculated emission spectrum. Moreover, this minimum becomes 
even slightly deeper if the short range part 
of this interaction in taken into account.  
Thus, the present model is unable to explain the appearance 
of the maximum at $\vartheta_k \simeq 90^0$ 
observed in the experimental data.   
 
\begin{figure}[t]
\vspace{0.25cm}
\begin{center}
\includegraphics[width=0.55\textwidth]{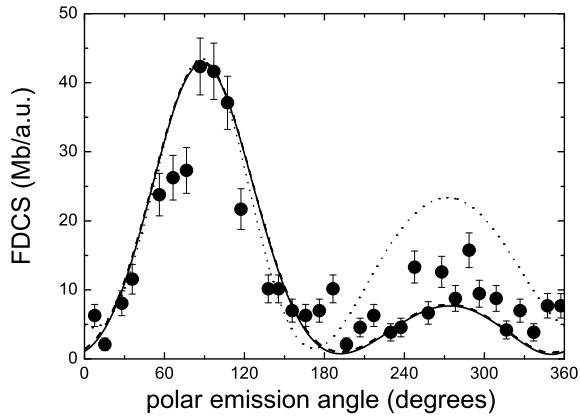}
\end{center}
\caption{ \footnotesize{ The FDCS for single
ionization of helium by $ 1 $ GeV/u U$^{92+}$ 
projectiles plotted as a function of the polar emission angle.
The cross section is given in the scattering plane. 
The electron emission energy $\varepsilon_k = 2 $ eV
and the total momentum transfer $ q = 0.65 $ a.u.. 
Symbols: experimental data from \cite{he-s-6}, 
these data have been fit to the maximum of the solid curve. 
Solid curve: results obtained by including into the interaction 
between the projectile and the target core 
the full expression (\ref{potential}). 
Dash curve: the projectile-target core interaction 
is modeled by the coulomb part of (\ref{potential}) only. 
Dot curve: the first Born results. }  }
\label{uranium-1} 
\end{figure}

\begin{figure}[t]
\vspace{0.25cm}
\begin{center}
\includegraphics[width=0.55\textwidth]{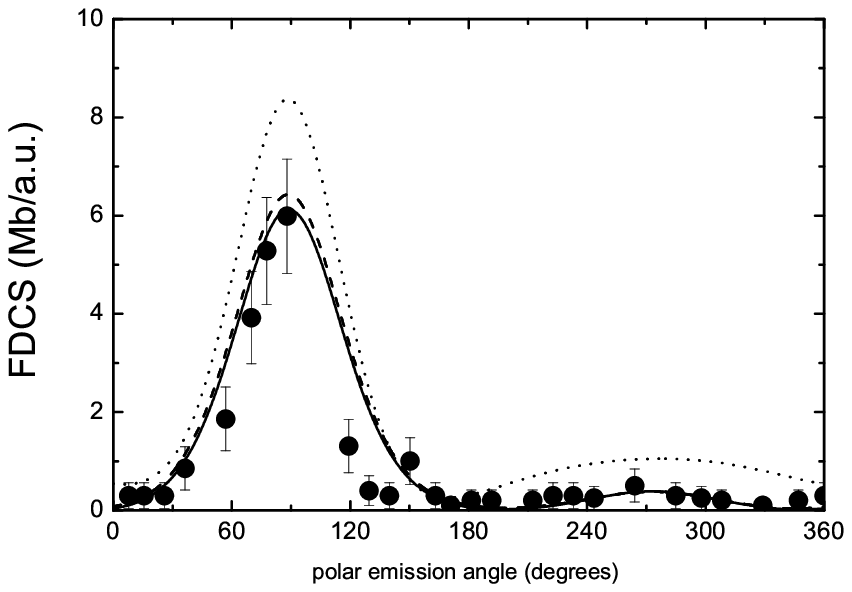}
\end{center}
\caption{ \footnotesize{ Same as in figure \ref{uranium-1} 
but for the electron emission energy of $10$ eV and 
the total momentum transfer of $1$ a.u.. }}
\label{uranium-2} 
\end{figure}

\begin{figure}[t]
\vspace{0.25cm}
\begin{center}
\includegraphics[width=0.55\textwidth]{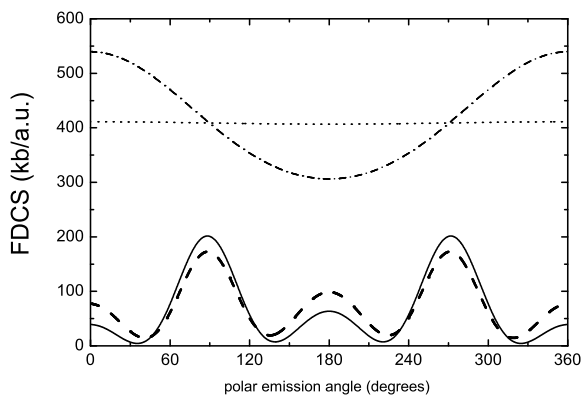}
\end{center}
\caption{ \footnotesize{ Same as in figure \ref{uranium-2} 
but the cross section is given in the perpendicular plane. 
Besides, the dash-dot curve displays results 
of the calculation in which the interaction between the projectile and 
the active electron is described within the rCDW-EIS but  
the projectile-target core interaction is ignored. }  }
\label{uranium-3} 
\end{figure}

Let us now turn to the FDCS in collisions with 
$1$ GeV/u U$^{92+}$ projectiles. 
In figures \ref{uranium-1}-\ref{uranium-2} we 
plot the FDCS for the scattering plane as a function 
of the polar angle of the emitted electron.
Theoretical results shown in these figures are on 
an absolute scale. The experimental 
data are normalized (at the cross section maximum) 
to those theoretical results 
which are obtained with the inclusion of 
the full interaction between the projectile 
and the target core. 

Like in the previous case, the main feature of 
the emission pattern is the presence 
of the binary and recoil peaks. 
However, in contrast to the case of helium ionization 
by $100$ MeV/u C$^{6+}$ ions, the calculated cross section 
in collisions with the $1$ GeV/u uranium ions 
substantially depends on 
whether or not the interaction between the projectile and 
the target core is taken into account. The effect of 
the short-range part of this interaction also becomes somewhat 
more visible. Nevertheless, for the collision parameters considered 
in figures \ref{uranium-1} and \ref{uranium-2} the calculated spectrum 
of the electron emission into the scattering plane still changes 
very little when the short-range part of the potential 
(\ref{potential}) is taken into account.   

Results for the electron emission into 
the perpendicular plane are shown in figure \ref{uranium-3}. 
Like in the case with the carbon projectiles, 
the calculated FDCS for the emission into this plane turns 
out to be much more sensitive to the approximations used. 
In particular, the inclusion of the interaction between 
the projectile and the target core substantially reduces  
the magnitude of the calculated cross section and 
has a very strong impact on its shape. Besides, 
the inclusion of the short range part of the potential 
(\ref{potential}) into this interaction results in a more pronounced 
effect. One more interesting feature of the emission pattern 
shown in figure \ref{uranium-3} is that the maximum in the emission 
is situated now at $\simeq 90^0$ and not at $0^0$ (or $180^0$) 
as it was in the case with the carbon projectiles. 
This finding is in agreement with our earlier 
predictions \cite{he-s-10},\cite{he-s-12} 
that, when the effective strength of the projectile field 
increases, a minimum in the calculated 
cross section at $\simeq 90^0$ disappears and, instead, 
the cross section attains a maximum at this point. 
  
\subsection{Cross section differential in the emission energy and 
transverse momentum transfer} 

\begin{figure}[t]
\vspace{0.25cm}
\begin{center}
\includegraphics[width=0.87\textwidth]{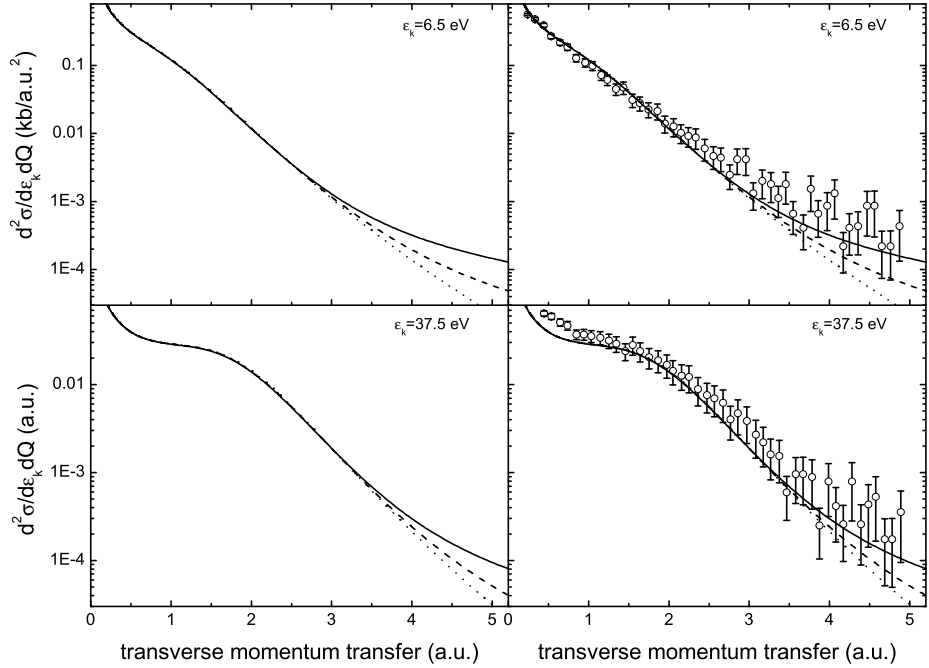}
\end{center}
\caption{ \footnotesize{ The doubly differential cross section 
$d^2 \sigma /d \varepsilon_k dQ$ for single
ionization of helium by $ 100 $ MeV/u C$^{6+}$ 
projectiles given as a function of the transverse momentum transfer 
$Q$ at a fixed value of the emission energy 
($\varepsilon_k = 6.5 $ and $\varepsilon_k = 37.5 $ eV 
for the upper and lower panels, respectively, of the figure).   
The left panels of the figure: solid curves are 
results obtained by including into the interaction 
between the projectile and the target core 
the full expression (\ref{potential});  
dash curves shows the results obtained 
when the projectile-target core interaction 
is modeled by the coulomb part of (\ref{potential}) only;  
dot curves are the first Born results. 
The right panels: the same as in the left panels plus 
experimental data from \cite{he-s-1}.}  }
\label{carbon-4} 
\end{figure}

\begin{figure}[t]
\vspace{0.25cm}
\begin{center}
\includegraphics[width=0.55\textwidth]{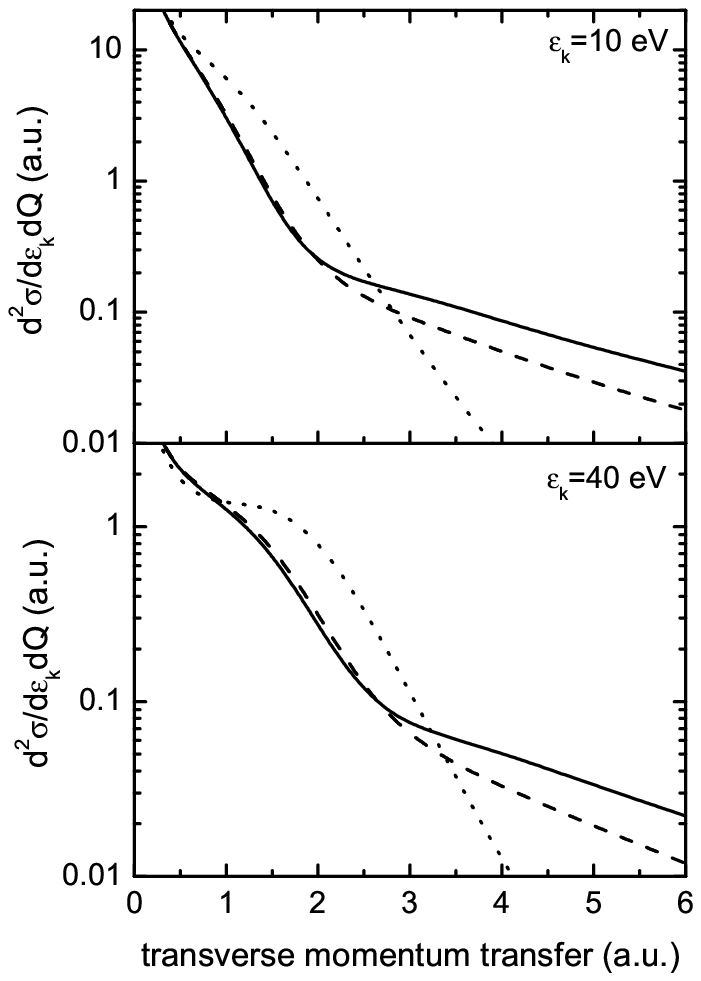}
\end{center}
\caption{ \footnotesize{ The doubly differential cross section 
$d^2 \sigma /d \varepsilon_k dQ$ for single
ionization of helium by $ 1 $ GeV/u U$^{92+}$ 
projectiles given as a function of the transverse momentum transfer 
$Q$ at a fixed value of the emission energy 
($\varepsilon_k = 10 $ and $\varepsilon_k = 40 $ eV 
for the upper and lower panels, respectively, of the figure).   
Solid curve: results obtained by including into the interaction 
between the projectile and the target core 
the full expression (\ref{potential}). 
Dash curve: the projectile-target core interaction 
is modeled by the coulomb part of (\ref{potential}) only. 
Dot curve: the first Born results.}  }
\label{uranium-4} 
\end{figure}

The influence of the interaction between 
the projectile and the target core on the dynamics of 
single ionization of helium in general depends not 
only on the electron emission energy, 
the momentum transfer and 
the magnitude of the parameter $\eta_p$ 
but also on a plane chosen to observe the FDCS.  
In order to get some more 'integral' ideas 
about the influence of the projectile-target core 
interaction as well as of the importance of 
its coulomb and short-range parts   
on the collision process we shall now consider  
the ionization cross section differential 
in the emission energy 
and the transverse part of the momentum transfer, 
$d^2 \sigma^{(+)}/dQ d\varepsilon_k$. 
Such a cross section is shown in figures 
\ref{carbon-4} and \ref{uranium-4} 
where it is considered as a function 
of the transverse momentum $Q$ at fixed values  
of the emission energy $\varepsilon_k$.   

In the case of single ionization of helium by 
$100$ MeV/u C$^{6+}$ projectiles noticeable differences 
in the cross section calculated without and with 
including the interaction between the projectile 
and the target core begin starting with 
$Q \simeq 3$-$4$ a.u. (see figure \ref{carbon-4}). 
In this figure we also see that at about the same values 
of $Q$ the short-range part of this interaction begins to have 
a visible effect on the cross section. This effect increases 
with increasing $Q$ and changes the cross section 
by about a factor of $2$ at $Q \simeq 5$ a.u..  
 
Compared to collisions with the carbon ions, 
the dynamics of single ionization of helium 
by $1$ GeV/u U$^{92+}$ projectiles 
is much stronger affected by the 
interaction between the projectile and the target core which 
is clearly seen in figure \ref{uranium-4}. 
Now the differences in the calculated cross sections 
caused by this interaction start already at $Q \simeq 0.5$ a.u.. 
Besides, the effect of the short range part of this interaction 
becomes noticeable at $Q \simeq 2$-$3$ a.u. and, thus, 
also comes into play at comparatively smaller values of $Q$.  

\section{Conclusions} 

We have considered some aspects of single ionization 
of helium by very fast highly charged nuclei. 
In this consideration we concentrated on the effects 
on the dynamics of this process 
caused by the interaction between 
the projectile-nuclei and the helium core and  
were especially interested in the role played 
by the short-range part of this interaction. 
As one could expect the role of this interaction 
as well as that of its short range part   
were found to increase when the transverse momentum 
transfer increases. It was also shown that they 
may result in very substantial effects. 
In particular, the inclusion of this interaction in the form 
(\ref{potential}) enables one to get a better description 
of the experimental data of \cite{he-s-1} 
on the doubly differential ionization 
cross section at larger values of $Q$. 

However, at smaller values of $Q$, 
for which the experimental data on 
the fully differential cross sections are available,   
our consideration shows that 
the short-range part of this interaction,  
both in collisions with $100$ MeV/u C$^{6+}$ 
and $1$ GeV/u U$^{92+}$ projectiles,  
has a very weak influence on 
the form and absolute values 
of the fully and doubly differential cross sections. 
Thus, our present study does not support 
the idea expressed in \cite{he-s-11} and \cite{he-s-11a}   
that the neglect of the short-range 
part of this interaction in the previous theoretical 
models could be responsible for the very substantial 
deviations between the experiment and theory in the case of 
the electron emission into the perpendicular plane.


\begin{thebibliography}{99}

\bibitem{1} J.Ullrich and V.P.Shevelko (ed), 
{\it Many-Particle Quantum Dynamics
in Atomic and Molecular Fragmentation}
(Springer-Verlag, 2003)  

\bibitem{he-s-1} D.Madison et al,  
J.Phys. {\bf B 35} 3297 (2002) 

\bibitem{he-s-2} 
M.Schulz et al. Nature 422, p. 48 (2003) 

\bibitem{he-s-3} D.H.Madison et al, 
Phys. Rev. Lett. {\bf 91} 253201 (2003) 

\bibitem{he-s-4} D.Fischer et al,
Phys. Rev. {\bf A 68} 032709 (2003)

\bibitem{he-s-5} M.Foster et al, J.Phys. {\bf B 37} 
1565 (2004)  

\bibitem{he-s-6} A.B.Voitkiv, B.Najjari, R.Moshammer, 
M.Schulz and J.Ullrich, J.Phys. {\bf B 37} L365 (2004)   

\bibitem{he-s-7} A.B.Voitkiv and B.Najjari, 
J.Phys. {\bf B 37} 4831 (2004) 

\bibitem{he-s-8} J.Fiol, S.Otranto and R.E.Olson,  
J.Phys. {\bf B 39} L285 (2006) 

\bibitem{he-s-9} M.D\"urr et al,  
Phys.Rev. {\bf A 75} 062708 (2007) 
 
\bibitem{he-s-10} A.B.Voitkiv, B.Najjari and J.Ullrich,  
J.Phys. {\bf B 36} 2591 (2003)   

\bibitem{he-s-11} M.Foster et al, 
Phys.Rev.Lett. {\bf 97} 093202 (2006) 

\bibitem{he-s-11a} M.Schulz et al, 
Phys.Rev. {\bf A 76} 032712 (2007) 

\bibitem{hartree-fock} F.Martin and A.Salin, Phys.Rev. {\bf A 55} 
2004 (1997) 

\bibitem{he-s-12} A.B.Voitkiv, B.Najjari and J.Ullrich,  
J.Phys. {\bf B 36} 2325 (2003)    

\end{thebibliography}
\end{document}